\documentclass[twocolumn,prl,preprintnumbers,amsmath,amssymb,letterpaper,superscriptaddress]{revtex4}

\newcommand{\m}[1]{\mathrm{#1}}

\usepackage[utf8]{inputenc}
\usepackage[T1]{fontenc}
\usepackage{graphicx}
\usepackage{dcolumn}
\usepackage{bm}
\usepackage{times}
\usepackage{amsmath}
\usepackage{amsfonts}
\usepackage{amssymb}
\usepackage{textcomp}

\usepackage{color}
\definecolor{red}{rgb}{1,0,0}
\definecolor{blue}{rgb}{0,0,1}

\newcommand{\X}{\mathrm{X}}
\newcommand{\IX}{\mathrm{IX}}
\newcommand{\XK}{\mathrm{X_K}}
\newcommand{\IXK}{\mathrm{IX_K}}

\begin{document}

\title{Cavity-control of bright and dark interlayer excitons in van der Waals heterostructures}

\author{Michael F\"org}
\affiliation{Fakult\"at f\"ur Physik, Munich Quantum Center, and
Center for NanoScience (CeNS), Ludwig-Maximilians-Universit\"at
M\"unchen, Geschwister-Scholl-Platz 1, D-80539 M\"unchen, Germany}
\author{L\'{e}o Colombier}
\affiliation{Fakult\"at f\"ur Physik, Munich Quantum Center, and
Center for NanoScience (CeNS), Ludwig-Maximilians-Universit\"at
M\"unchen, Geschwister-Scholl-Platz 1, D-80539 M\"unchen, Germany}
\author{Robin K. Patel}
\affiliation{Fakult\"at f\"ur Physik, Munich Quantum Center, and
Center for NanoScience (CeNS), Ludwig-Maximilians-Universit\"at
M\"unchen, Geschwister-Scholl-Platz 1, D-80539 M\"unchen, Germany}
\author{Jessica Lindlau}
\affiliation{Fakult\"at f\"ur Physik, Munich Quantum Center, and
Center for NanoScience (CeNS), Ludwig-Maximilians-Universit\"at
M\"unchen, Geschwister-Scholl-Platz 1, D-80539 M\"unchen, Germany}
\author{Aditya D. Mohite }
\affiliation{MPA-11 Materials Synthesis and
Integrated Devices, Materials Physics and Applications Division, Los Alamos National Laboratory (LANL), Los Alamos, New Mexico 87545, USA}
\author{Hisato Yamaguchi}
\affiliation{MPA-11 Materials Synthesis and
Integrated Devices, Materials Physics and Applications Division, Los Alamos National Laboratory (LANL), Los Alamos, New Mexico 87545, USA}
\author{David Hunger}
\affiliation{Physikalisches Institut, Karlsruher Institut f\"ur Technologie, Wolfgang-Gaede-Straße 1, 76131 Karlsruhe, Germany}
\author{Alexander H\"ogele}
\affiliation{Fakult\"at f\"ur Physik, Munich Quantum Center, and
Center for NanoScience (CeNS), Ludwig-Maximilians-Universit\"at
M\"unchen, Geschwister-Scholl-Platz 1, D-80539 M\"unchen, Germany}

\date{\today}

\begin{abstract}
Monolayer (ML) transition metal dichalcogenides (TMDs) integrated
in optical microcavities host exciton-polaritons as a hallmark of
the strong light-matter coupling regime. Analogous concepts for
hybrid light-matter systems employing spatially indirect excitons
with a permanent electric dipole moment in heterobilayer (HBL)
crystals promise realizations of exciton-polariton gases and
condensates with immanent dipolar interactions. Here, we identify
optical signatures of spatially indirect momentum-bright and
momentum-dark interlayer excitons in vertical MoSe$_2$-WSe$_2$
heterostructures and implement cavity-control of both exciton
manifolds. Our experiments quantify the strength of light-matter
coupling for both zero and finite momentum excitons residing in
Moir\'{e} superlattices of TMD HBLs and demonstrate that both
exciton species are susceptible to Purcell enhancement in
cavity-modified photonic environments. Our results form the basis
for further developments of dipolar exciton-polariton gases and
condensates in hybrid cavity -- van der Waals heterostructure
systems.
\end{abstract}

\maketitle


Semiconductor TMDs exhibit remarkable optoelectronic and
valleytronic properties in the limit of direct band-gap MLs
\cite{Splendiani2010,Mak2010,Xiao2012,Xu2014}. High oscillator
strength renders the materials ideal for the studies of collective
strong coupling phenomena mediated among excitons and photons by
optical resonators. This limit of new bosonic eigenstates of
half-matter and half-light quasiparticles known as
exciton-polaritons was first observed for ML TMDs with MoS$_2$ in
a planar microcavity at room temperature \cite{Liu2015}. Subsequent studies
established enhanced coupling rates for cavity-integrated WS$_2$
MLs \cite{Flatten2016}. Combinations of dielectric and metallic
mirrors allowed the observation of Tamm-plasmon polaritons
\cite{Lundt2016} with substantial degrees of valley polarization
\cite{Lundt2017}, while additional control of the TMD doping level
has enabled studies of fermionic polaron-polaritons
\cite{Sidler2017}.

In contrast to TMD MLs, cavity-control of their heterobilayer
(HBL) counterparts has been elusive despite their potential for
fundamental studies of dipolar gases with intriguing polarization
dynamics upon expansion \cite{Rivera2016} and condensation
phenomena \cite{Fogler2014}. Composed of two dissimilar MLs in
staggered band alignment \cite{Kang2013,Chiu2015}, such van der
Waals heterostructures host layer-separated electron-hole pairs in
response to optical excitation \cite{Rivera2015}. The spatial
separation of Coulomb-correlated electrons and holes gives rise to
a permanent exciton dipole moment along the stacking axis and
extended lifetimes up to hundreds of ns
\cite{Rivera2015,Miller2017,Jiang2017,Korn2017}. While long
lifetimes are beneficial in providing sufficient time for
thermalization, finite exciton dipole moments underpin mutual
interactions in exciton-polaritons gases and condensates. To date,
however, the integration of HBLs into optical cavities has been
impeded by the involved fabrication method of exfoliation-stacking
which requires careful alignment of both MLs along the
crystallographic axes to reduce momentum mismatch between
electrons and holes residing in dissimilar layers.

As opposed to exfoliation-stacking, chemical vapor deposition
(CVD) realizes inherently aligned TMD heterostructures with
atomically sharp interfaces both in lateral and vertical
geometries \cite{Gong2014,Li2015}. Such CVD-grown vertical
MoSe$_2$-WSe$_2$ heterostructures were used in our experiments to
demonstrate cavity-control of interlayer excitons for the first
time. To this end we employed a tunable open-access cavity with
one curved fiber-based mirror and one planar mirror with extended
heterostructure flakes on top (see Supplementary Information for
details on sample fabrication). The configuration of controlled
intermirror spacing and lateral scanning capabilities with proven
potential in the studies of other low-dimensional condensed matter
systems \cite{Hummer2016,Kaupp2016,Benedikter2017} was used to
explore the light-matter coupling of excitons in an extended
MoSe$_2$-WSe$_2$ heterostructure as a function of the cavity
length at representative positions of HBLs selected by
two-dimensional cavity imaging.

\begin{figure}[t]
\includegraphics[scale=1]{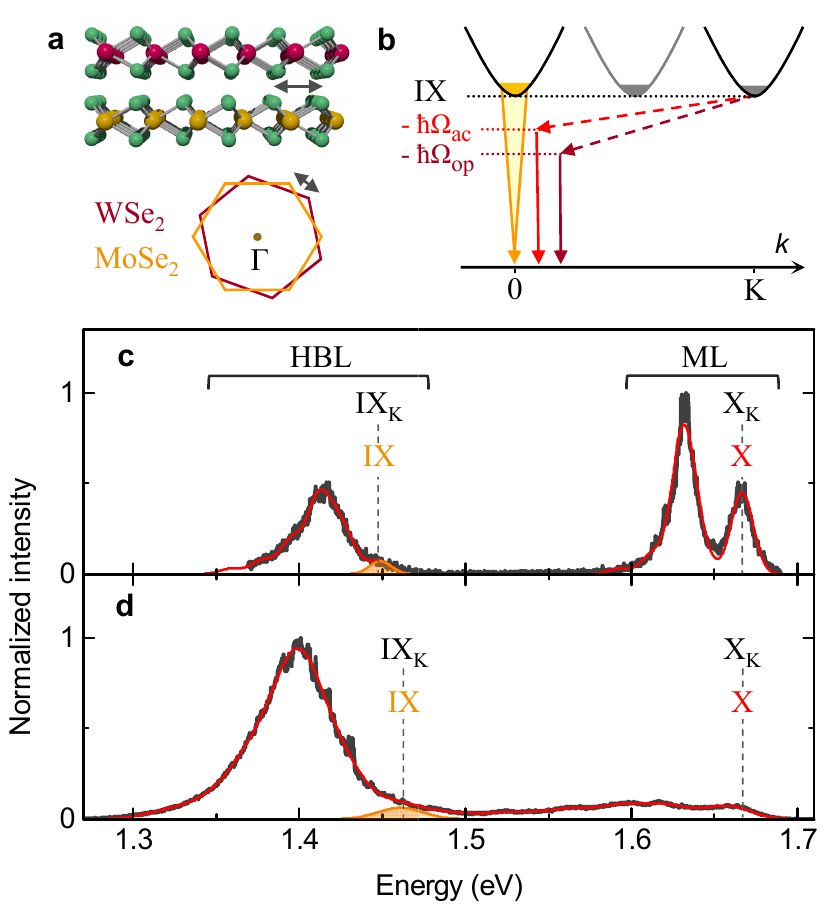}
\caption{\textbf{a}, Real space and reciprocal space schematics of
a MoSe$_{2}$-WSe$_{2}$ heterostructure with lattice mismatch
symbolized by the arrow. \textbf{b}, Momentum-direct excitons
(orange-shaded population) within the light cone decay radiatively
with interlayer exciton PL energy $\IX$. Momentum-dark excitons
(grey-shaded populations away from the zero-momentum origin) decay
only with the assistance of phonons and thus contribute to PL as
acoustic and optical phonon sidebands downshifted from $\IX$ by
$\m{\hbar\Omega_{ac}}$ and $\m{\hbar\Omega_{op}}$, respectively.
\textbf{c}, Spectral decomposition (red solid line) of the
cryogenic PL from an as-grown MoSe$_{2}$-WSe$_{2}$ sample into the
contributions of radiative intralayer and interlayer excitons at
energy $\X = 1.67$~eV and $\IX = 1.45$~eV, respectively, and
phonon sidebands of momentum-dark intralayer and interlayer
excitons (with energies $\XK$ and $\IXK$ resonant with $\X$ and
$\IX$, respectively, as indicated by the dashed lines). The zero
phonon line of momentum-bright interlayer excitons is explicitly
shown by the orange-shaded Gaussian with an inhomogeneous
linewidth of $\gamma = 16$~meV obtained from best-fit. \textbf{d},
Same but for a MoSe$_{2}$-WSe$_{2}$ heterostructure on the cavity
mirror. Best-fit energies of the intralayer and interlayer exciton
energies (shown by the dashed lines) were $\X = 1.66$~eV and $\IX
= 1.46$~eV, respectively, with with an inhomogeneous linewidth of
$\gamma = 32$~meV consistent with higher disorder than in the
as-grown sample of \textbf{c}.} \label{fig1}
\end{figure}

Before demonstrating cavity-control of excitons in a HBL system we
identified the main signatures of intralayer and interlayer
excitons with cryogenic photoluminescence (PL) spectroscopy. The
crystal schematic of the van der Waals heterostructure,
synthesized by overgrowth of ML MoSe$_2$ with a ML of WSe$_2$, is
shown in Fig.~\ref{fig1}a. Upon optical excitation,
photo-generated electrons and holes relax into the conduction band
minima and valence band maxima of MoSe$_2$ and WSe$_2$ MLs,
respectively \cite{Rivera2015}. After relaxation, zero-momentum
excitons are formed by layer-separated electrons and holes of the
same valley, while excitons with finite momentum emerge from
Coulomb correlations among electrons and holes in dissimilar
valleys. Even in the presence of inherent angular alignment,
excitons in HBL stacks exhibit a large spread in their
center-of-mass momenta just like twisted HBL systems
\cite{Yu2015}. This spread in momentum stems from Moir\'e
superlattices \cite{Tong2016} which in turn are a consequence of
the lattice mismatch (symbolized by the arrows in
Fig.~\ref{fig1}a) in HBL systems built from dissimilar ML
crystals. Thus, only the small fraction of zero-momentum
interlayer excitons (labelled as IX in Fig.~\ref{fig1}b) decays
within the radiative lifetime of a few ns \cite{Rivera2015}. On
the same timescale, a significant population of much longer lived
excitons with center-of-mass momentum beyond the light cone (with
exciton dispersions in Fig.~\ref{fig1}b shifted away from the
origin up to the maximum in-plane momentum K) will accumulate due
to highly suppressed radiative decay.

In confocal PL spectroscopy, diffraction-limited excitation and
detection spots sample a large number of Moir\'e superlattice
cells and thus probe both momentum-bright and momentum-dark
interlayer excitons. The former emit PL directly at their bare
energy of zero-momentum interlayer excitons $\IX$. The latter, on
the other hand, contribute to the PL spectrum as phonon sidebands
\cite{Lindlau2017ML,Lindlau2017BL} downshifted from $\IX$ by the
energy of acoustic or optical phonons ($\m{\hbar\Omega_{ac}}$ or
$\m{\hbar\Omega_{op}}$) that compensate for momentum-mismatch in
the light-matter coupling and thus enable radiative decay of
momentum-dark excitons. Both the zero phonon line (ZPL) of
momentum-bright excitons $\IX$ and the phonon sidebands of
momentum-dark excitons $\IXK$ contribute to the intense red-most
peak around $1.4$~eV in the PL spectra of our as-grown and
mirror-transferred MoSe$_2$-WSe$_2$ HBLs shown in Fig.~\ref{fig1}c
and d, respectively. In addition to this low-energy HBL peak
arising from interlayer excitons as in exfoliation-stacked
heterostructures \cite{Rivera2015,Miller2017,Jiang2017,Korn2017},
intralayer MoSe$_2$ excitons contribute a pair of blue peaks to
the PL around $1.65$~eV \cite{Rivera2015}. Commonly, the upper and
lower ML peaks are assigned to neutral and charged excitons
(trions) in MoSe$_2$ \cite{Ross2013}, respectively.

In the presence of a long-lived reservoir of momentum-dark
counterparts ($\XK$) of momentum-bright neutral ML excitons
($\X$), however, this assignment is ambiguous. In fact, the trion
signatures as well as the extended red tail in the PL below $\X$
can be explained on the basis of phonon replicas of momentum-dark
excitons alone \cite{Lindlau2017ML}. We used this notion to
decompose the PL spectra of Fig.~\ref{fig1}c and d into
contributions from momentum-bright and -dark intralayer ($\X$ and
$\XK$) and interlayer ($\IX$ and $\IXK$) excitons with the energy
positions of their respective ZPL identified from best-fit (red
solid lines) and indicated by the dashed lines in Fig.~\ref{fig1}c
and d. The states $\X$ and $\XK$ as well as $\IX$ and $\IXK$ were
assumed energy-degenerate by neglecting electron-hole exchange and
modelled by Gaussian ZPLs with identical inhomogeneous linewidth
$\gamma$. The differences in the absolute energies of $\X$ and
$\IX$ as well as the difference in spectral broadening in the two
samples probably stem from different dielectric environments,
strain and other disorder. More remarkably, our spectral
decomposition analysis, analogous to recent studies of
complementary TMD systems \cite{Lindlau2017ML,Lindlau2017BL}, not
only explains the PL between $\X$ and $\IX$ in terms of
higher-order phonon replicas activated by disorder
\cite{Lindlau2017ML}, it also implies that only a small fraction
of the characteristic HBL peak actually stems from radiative
interlayer excitons $\IX$ (with ZPLs given explicitly by the
yellow-shaded Gaussians in Fig.~\ref{fig1}c and d). The
predominant part of the HBL peak is ascribed to acoustic and
optical phonon sidebands of the momentum-dark exciton reservoir at
$\IXK$.

\begin{figure}[t]
\includegraphics[scale=1]{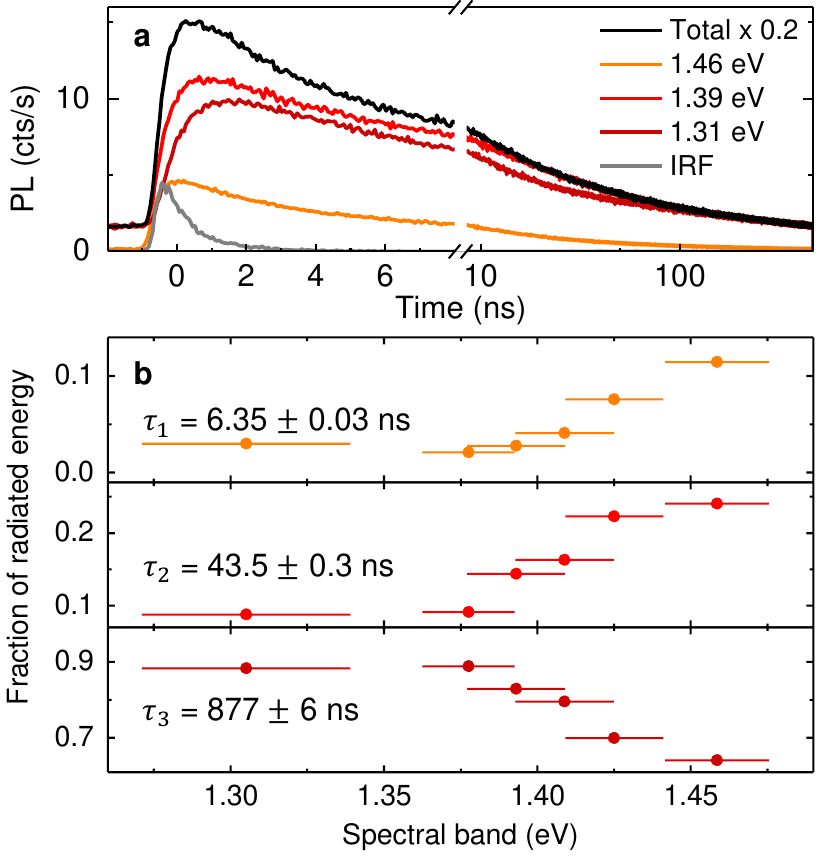}
\caption{\textbf{a}, PL decay measured in different spectral bands
on a mirror-transferred MoSe$_{2}$-WSe$_{2}$ flake with the
spectrum in Fig.~\ref{fig1}d. Decay traces are shown in different
colors for three spectral bands centered at $1.46$, $1.39$ and
$1.31$~eV together with the total decay trace in black (scaled by
$\times 0.2$) and the instrument response function (IRF) in grey.
The decay of the total spectrally unfiltered PL was approximated
best by three exponential decay channels with constants of $6$,
$43$ and $877$~ns. \textbf{b}, Relative contributions of the three
decay channels to the total PL radiated into the spectral windows
with central energies and widths represented by dots and bars,
respectively. For each spectral window, the contributions were
extracted from triple-exponential fits with decay constants fixed
to the characteristic timescales of the total PL.} \label{fig2}
\end{figure}

In order to substantiate this new perspective on the PL signatures
of HBLs we carried out time-resolved PL experiments on a
mirror-transferred MoSe$_2$-WSe$_2$ flake. Previous cryogenic
studies of exfoliation-stacked MoSe$_2$-WSe$_2$ heterostructures
reported a broad range of phenomena for the dynamics of interlayer
exciton PL with lifetimes in the range of $1 - 100$~ns extracted
from single- or multi-exponential decay fits
\cite{Rivera2015,Rivera2016,Miller2017,Jiang2017,Korn2017}. The
broad-band interlayer HBL peak of our CVD-grown samples exhibits
similar decay characteristics. The best approximation to the total
HBL peak was obtained with three exponential decay channels
characterized by lifetimes of $\sim 6$, $44$ and $877$~ns (see
Supplementary Information for details). Interestingly, the
contributions of the individual decay channels to the total
radiated PL energy varied significantly across the HBL peak. By
performing PL decay measurements in narrow spectral windows
centered at variable energies as in Fig.~\ref{fig2}a, we found
that the relative weight of the slowest decay component with
$877$~ns decay constant increased at the expense of the more rapid
components with $6$ and $44$~ns lifetimes as the spectral band of
the measurement window was shifted to lower energies
(Fig.~\ref{fig2}b). In the red-most wing, interlayer PL was
significantly delayed (note the prolonged rise-time of the PL
traces in Fig.~\ref{fig2}a recorded in the red wing) and nearly
entirely dominated by the longest decay constant. Its finite value
on the blue side of the peak is attributed to disorder-mediated
slow multi-phonon decay channels \cite{Cassabois2016a} of
intralayer excitons $\X$.

The cross-over from short to long PL lifetimes upon progressive
red-shift is consistent with our interpretation of the total HBL
PL. Our model predicts a decrease for the PL contribution of the
momentum-bright exciton state $\IX$ upon increasing red-shift from
its ZPL, and this trend is consistently supported by the data in
the upper panel of Fig.~\ref{fig2}b. The secondary lifetime due to
decay processes of momentum-dark interlayer excitons $\IXK$
assisted by first-order acoustic and optical phonons exhibits the
same trend yet at larger red-shifts (data in the central panel of
Fig.~\ref{fig2}b). Finally, the contribution of the third decay
component stemming from higher-order decay processes assisted by
multiple phonons is suppressed at small red-shifts and dominates
only for largest red-shifts (data in the lower panel of
Fig.~\ref{fig2}b). Before proceeding we note that this framework
of three characteristic decay components might be too rudimentary
to capture the actual population dynamics of interlayer excitons
in HBLs with a variety of decay channels associated with the
spread in the exciton momenta by the Moir\'{e} effects and
additional disorder (as evidenced by the PL in between $\X$ and
$\IX$ peaks in Fig.~\ref{fig1}d). However, it is also not obvious
how to introduce three specific decay timescales in the framework
of disorder-localized excitons commonly employed to motivate
long-lived PL decays.


\begin{figure}[t]
\includegraphics[scale=1]{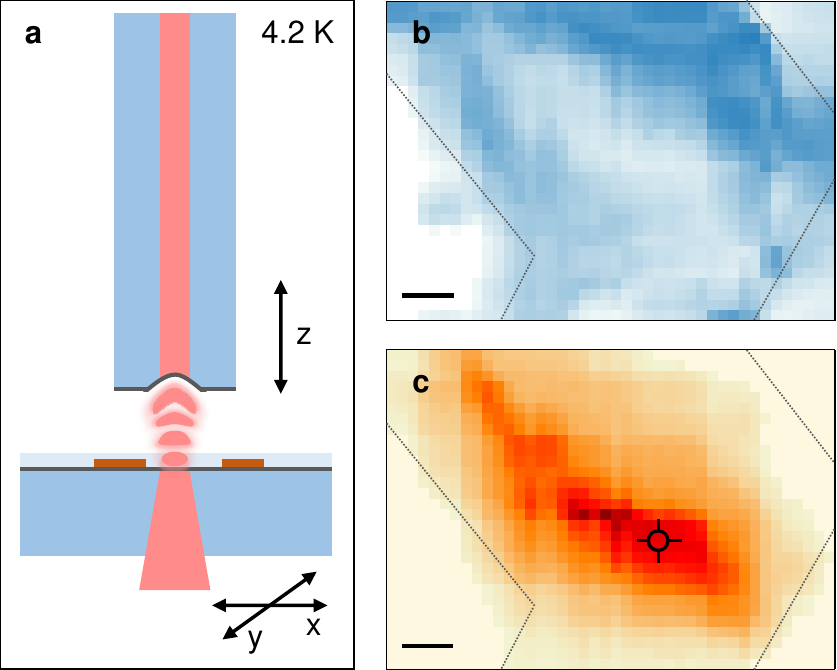}
\caption{\textbf{a}, Cavity setup at $4.2$~K: the fiber-based
micro-mirror forms the cavity together with a planar macro-mirror
with CVD-grown MoSe$_{2}$-WSe$_{2}$ on top. Independent
translational degrees of freedom enable lateral sample
displacement and cavity length detuning. \textbf{b}, Transmission
map recorded through the cavity with laser excitation at $635$~nm
(blue color corresponds to reduced transmission due to local
variations in absorption and scattering). \textbf{c}, Map of
integrated PL intensity recorded simultaneously with the
transmission map (dark red color represents maximum intensity).
The cross indicates the position on the flake used in the
measurements of Fig.~\ref{fig2} and Fig.~\ref{fig4}, the grey dashed
lines indicate the boundaries of the flake. The scale bar
is $10~\mu$m in both maps.} \label{fig3}
\end{figure}

Bearing this limitation in mind we proceed by presenting results
on cavity-control of the three characteristic PL decay channels.
To this end, we complemented the macroscopic mirror with CVD-grown
MoSe$_2$-WSe$_2$ flakes on top by a fiber micro-mirror in a
cryogenic setup (see Supplementary Information for details). In
brief, our optical Fabry-P\'erot resonator, immersed in a helium
bath cryostat at $4.2$~K, is based on a single-mode fiber with a
laser-machined concave end facet coated with silver and a
protecting layer of SiO$_2$. The schematic drawing of the cavity
setup with independent translational degrees of freedom along all
three dimensions is shown in Fig.~\ref{fig3}a. Lateral
displacement of the sample mirror enabled coarse-tuning of the
cavity length as well as two-dimensional positioning and profiling
of the sample. The respective transmission and PL maps of the
flake with PL data in Fig.~\ref{fig1}d and Fig.~\ref{fig2} are
shown in Fig.~\ref{fig3}b and c.

The transmission of the excitation laser at $635$~nm provided
quantitative access to absorption and scattering inside the
cavity. Sizeable ML absorption in the range of several percent
\cite{Li2014} facilitated the detection of individual MLs and HBL
via the cavity transmission. Scattering contrast at structural
defects such as edges or transfer-related cracks provided
additional guides to the identification of individual flakes.
Equipped with the combined scanning capabilities and the
information from transmission, it was straight forward to position
the cavity into any point of interest of a given HBL flake. In
addition, by recording PL spectra at each raster scan-point of the
cavity simultaneously with the transmission, PL intensity maps
were obtained within any spectral band of interest as shown for
the interlayer exciton PL peak in Fig.~\ref{fig3}c.

\begin{figure}[t]
\includegraphics[scale=1.0]{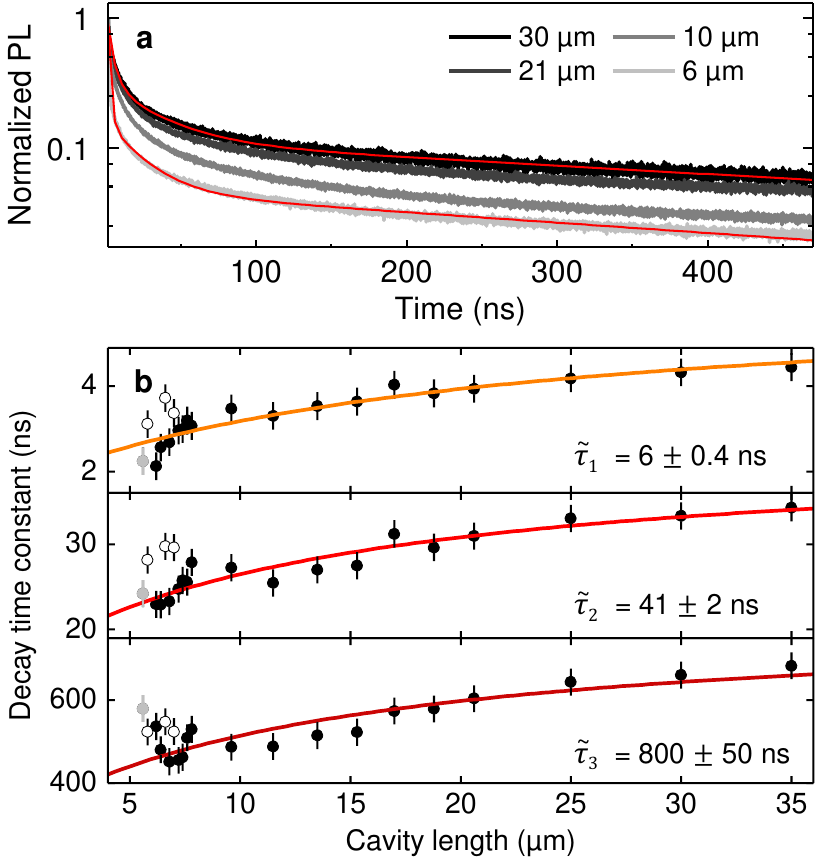}
\caption{\textbf{a}, Traces of interlayer exciton PL decay shown
for four selected cavity lengths.  The solid lines are fits to the
data with three exponential decay constants. Note the speed up in
the decay upon the reduction of the cavity length. \textbf{b},
Evolution of the characteristic decay constants with the cavity
length. The solid lines show model fits according to the theory of
generalized Purcell enhancement. Open circles represent data where
the cavity mode was spectrally detuned from the resonance with the
interlayer peak; data shown in light grey were discarded from the
fit procedure due to presumable physical contact between the fiber
and the mirror.} \label{fig4}
\end{figure}

By monitoring both transmission and PL, we positioned the cavity
on the spot indicated by the cross in Fig.~\ref{fig3}c where data
of Fig.~\ref{fig1}d and Fig.~\ref{fig2} were recorded with
confocal spectroscopy and performed PL decay measurements as a
function of the cavity length. The respective decay traces of the
HBL peak in the spectrum of Fig.~\ref{fig1}d are shown in
Fig.~\ref{fig4}a for cavity lengths of $30$, $21$, $10$ and
$6~\mu$m. Clearly, the PL decay speeds up with decreasing cavity
length. For a more quantitative analysis, the PL traces recorded
at different cavity lengths were modeled by a convolution of the
instrument response function (IRF) and a three-exponential decay
with amplitudes and time constants of each decay channel as free
fit parameters (see Supplementary Information for details). The
corresponding model fits, shown as red solid lines in
Fig.~\ref{fig4}a, were used to extract the evolution of the short,
intermediate and long decay time components with the cavity
length.

The respective data, shown in Fig.~\ref{fig4}b, clearly
demonstrate cavity-control of all three characteristic decay
channels. The lifetimes decrease with decreasing cavity length - a
hallmark of Purcell enhancement which can be quantified by the
ratio of the total decay rate in the cavity system $\gamma_c$ to
the free-space decay rate $\gamma_{fs}$ as $\gamma_c/\gamma_{fs} =
1 + C$, where $C$ is the Purcell factor. An estimate for the
cavity-mediated Purcell enhancement can be obtained by identifying
the values from confocal decay measurements with free-space
lifetimes. Taking the smallest lifetime values for each decay
channel from the data of Fig.~\ref{fig4}b, this yields maximum
measured Purcell factors of $1.8 \pm 0.3$, $0.8 \pm 0.1$ and $0.9
\pm 0.1$ for the short, intermediate and long lifetime components,
respectively. The difference in the Purcell factors highlights the
different nature of the coupling between the corresponding decay
channels and the cavity field, with interlayer momentum-bright
excitons $\IX$ exhibiting higher coupling efficiency than
phonon-mediated decay channels of momentum-dark excitons $\IXK$.

The three different channels also exhibit pronounced differences
in their response to cavity length detuning shown in
Fig.~\ref{fig4}b. At a cavity length of $35~\mu$m, several cavity
modes are resonant with the HBL emission peak thus enhancing all
possible emission channels simultaneously. For cavity lengths
smaller than $9~\mu$m, however, the free spectral range of the
cavity exceeded the linewidth of the HBL emission peak, rendering
light-matter coupling sensitive to the spectral resonance
condition. Open circles in Fig.~\ref{fig4}b show the results for
the off-resonant configurations in accord with cavity-inhibited
radiative decay. In contrast, the on-resonance data (measured with
a dense spacing of data points for $\sim 6 - 8~\mu$m cavity
lengths in Fig.~\ref{fig4}b) reflect the effect of
cavity-enhancement with anti-correlated trends for short and long
decay components at smallest cavity lengths consistent with
spectrally distinct channels. At a nominal separation of $\sim
5~\mu$m (grey circles), physical contact between the fiber and the
extended mirror was presumably reached, preventing further
reduction of the cavity mode volume.

The data points recorded in contact of the fiber and the
macro-mirror as well as all off-resonance data were discarded from
the following analysis of the cavity-induced Purcell enhancement
in the presence of pure dephasing \cite{Auffeves2010}. On
resonance, the generalized Purcell factor is $C =
(4g^2/\gamma_{fs})/(\kappa +\gamma_{fs} + \gamma_d)$, where $g$ is
the coupling rate of the emitter to the cavity, $\kappa$ is the
cavity decay rate, and $\gamma_d$ is the dephasing rate of the
emitter. Both $g$ and $\kappa$ vary as a function of the cavity
length \cite{Savona1995,Besga2015,Hunger2010}. By taking the
inhomogeneous linewidth $\gamma=32$~meV obtained from model fits
of Fig.~\ref{fig1}d as an upper bound to the dephasing rate in our
system (i. e. using $\gamma_d=\gamma$), we fitted each data set of
Fig.~\ref{fig4}b according to the model for the generalized
Purcell enhancement (see Supplementary Information for details).
The resulting best fits, shown as solid lines in Fig.~\ref{fig4}b,
were obtained with free-space lifetimes of $6.2 \pm 0.4$, $41 \pm
2$ and $800 \pm 50$~ns for the three sets of data in the
respective panels of Fig.~\ref{fig4}b. These asymptotic values at
infinite cavity length extracted from the model fit agree well
with the decay times determined in confocal PL spectroscopy (data
in Fig.~\ref{fig2}b).

With this strong confidence in the correspondence between the
free-space lifetime values extracted from the model of generalized
Purcell enhancement and the decay times obtained in the absence of
the cavity with confocal PL spectroscopy, the model allows now to
extrapolate maximum Purcell enhancement $C^{\m{max}}$ that can be
achieved at the peak wavelength of the HBL emission $\lambda$ for
a mirror separation of $\lambda/2$. The model yields $C^{\m{max}}$
of $2.5\pm 0.2$ for the short and $ 1.4\pm 0.1$ for both the
intermediate and long lifetime channels, respectively. For the
same limit of the intermirror spacing of $\lambda/2$ and a cavity
volume of $\sim \lambda^3$, the model also quantifies the
light-matter coupling strength $g$ as $170 \pm 8$, $50 \pm 2$ and
$11 \pm 0.7~\mu$eV for the short, intermediate and long decay
channel, respectively. These values are rather robust against
variations in the dephasing rate, with $g$ changing by less than
$25~\%$ for $\gamma_d$ in the range of $10 - 70$~meV. At the same
time light-matter coupling was sensitive to material and
environmental characteristics with up to $50~\%$ changes in $g$
and about $30~\%$ variations in the free-space PL lifetimes on
different positions of the same flake and different flakes.

The values for the light-matter coupling strength $g$ of
interlayer excitons in our CVD-grown MoSe$_2$-WSe$_2$ HBL sample
are two to three orders of magnitude smaller than the coupling
rates reported for MLs TMDs
\cite{Liu2015,Flatten2016,Lundt2017,Lundt2016}. This striking
difference in light-matter coupling, fully consistent with the
spatially indirect nature of momentum-bright and momentum dark
interlayer excitons in HBL systems, yields tight constraints on
the observation of interlayer exciton-polariton phenomena in the
strong-coupling regime of HBL -- cavity hybrids. To ensure $g >
\kappa + \gamma_d$ for strong-coupling, cavities with higher
quality factors are readily available \cite{Hummer2016}, yet much
improved HBL crystals and environmental conditions will be
required to reduce dephasing. However, in view of radiatively
limited linewidths achieved for ML TMDs by encapsulation with
hexagonal boron nitride \cite{Wang2017a,Cadiz2017,Ajayi2017},
further progress towards the realization of dipolar
exciton-polariton gases in cavity -- van der Waals heterostructure
systems seems feasible.

{\bf Acknowledgments:} We thank Guillaume Cassabois (Universit\'e
Montpellier) for fruitful discussions and acknowledge support from
Yongji Gong (Beihang University) and Pulickel M. Ajayan (Rice
University) on material growth parameters at the initial stage of
project. This research was funded by the European Research Council
under the ERC Grant Agreement no. 336749, the Volkswagen
Foundation, and the German Excellence Initiative via the
Nanosystems Initiative Munich (NIM). A.~H. also acknowledges
support from the Center for NanoScience (CeNS) and LMUinnovativ.
A.~D.~M. acknowledges support from LDRD program and CINT at LANL.

\setcounter{figure}{0} \setcounter{equation}{0}
\renewcommand{\figurename}{Figure~S}
\makeatletter
\def\fnum@figure{\figurename\thefigure}
\makeatother
\newcommand{\figref}[1]{Fig.~S\ref{#1}}

\section{Supplementary Information}

\subsection{Experimental methods}

\subsubsection{Chemical vapor deposition of TMD heterobilayers}

First, MoSe$_2$ ML was grown by selenization of molybdenum
trioxide (MoO$_3$) powder. SiO$_2$/Si substrate along with MoO$_3$
powder boat were placed at the center of a chemical vapor
deposition (CVD) furnace, which was heated to 750~°C in 15 min
and held for 20 min. SiO$_2$/Si substrate was facing down in close
proximity with MoO$_3$ powder. Selenium (Se) powder vaporized at
200~°C was used as Se source, and a mixture of argon and hydrogen
(15\% hydrogen) at 50 SCCM was used as the carrier gas. The
as-grown MoSe$_2$/SiO$_2$/Si was then transferred to a separate
CVD setup for subsequent WSe$_2$ growth similar to the growth
method of MoSe$_2$. Specifically, selenization of tungsten oxide
(WO$_3$) was performed at 900~°C in the presence of 100 SCCM
carrier gas. WSe$_2$ would grow on top of MoSe$_2$ from its edges,
creating MoSe$_2$/WSe$_2$ vertical heterostructures. No additional
treatment was necessary prior to WSe$_2$ growth due to thermal
removal of possible physisorbed molecule gases on MoSe$_2$ during
the transfer in air. As-grown heterostructures were studied in
spectroscopy or transferred onto a mirror using established
polymer-supported wet method. To this end poly(methyl
methacrylate) (PMMA) was spin-coated on the heterostructure and
lifted off in 1M possosium hydroxide (KOH) in water. Finally, the
PMMA-supported film with MoSe$_2$-WSe$_2$ vertical
heterostructures on the mirror was rinsed in three cycles of water
at room temperature to remove possible KOH residue.

\subsubsection{Photoluminescence microscopy and spectroscopy}

Photoluminescence experiments were performed in a home-built
cryogenic scanning cavity setup. The sample was mounted on
piezo-stepping units (attocube systems ANPxy101 and ANPz102) for
positioning with respect to the cavity mode. The cavity unit was
placed in a dewar with an inert helium atmosphere at a pressure of
$20$~mbar and immersed  in liquid helium at $4.2$~K. Excitation
around $635 - 705$~nm was performed with a wavelength-tunable
white light laser system (NKT SuperK Extreme and SuperK Varia)
with repetition rates down to 2~MHz. The PL was either spectrally
dispersed by a monochromator (Princeton Instruments Acton SP 2500)
and recorded with a nitrogen-cooled silicon CCD (Princeton
Instruments PyLoN) or detected with avalanche photodiodes
(Excelitas SPCM-AQRH or PicoQuant $\tau$SPAD).\\

\subsubsection{Scanning cavity microscopy}

The cavity was composed of a fiber micro-mirror and a macroscopic
mirror with MoSe$_2$-WSe$_2$ vertical HBL on top. The macro-mirror
was coated with $\sim 30~$nm of silver and a spacer layer of
SiO$_2$ with thickness designed to place the HBL at a field
antinode. The effective radius of curvature of the central
depression in the laser-machined fiber end facet was $136~\mu$m.
The facet was coated with $\sim 50~$nm silver and a protection
layer of SiO$_2$. Three translational degrees of freedom of the
sample on the mirror was enabled by cryogenic piezo-stepping units
(attocube systems ANPxy101 and ANPz102) to provide both lateral
scans and coarse-tuning of the cavity length. Cavity fine-tuning
was achieved by displacing the fiber-mirror with an additional
piezo. Excitation by a diode laser at $635$~nm was provided via
the optical fiber and both transmission and PL were detected
through the planar macro-mirror with the heterostructure on top.
Two-dimensional scans were performed with a cavity length of $\sim
22~\mu$m resulting in a mode-waist of $3.2~\mu$m for the
excitation laser and a mode-waist of $3.7~\mu$m for the detected
PL around $880$~nm.

\subsection{Analysis of time-correlated photoluminescence decay}\label{fitting}

\subsubsection{Deconvolution procedure}

\begin{figure}[h]
\includegraphics[scale=1]{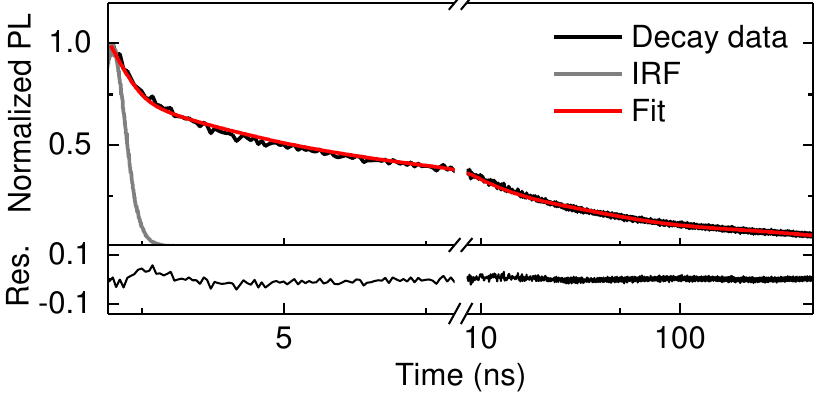}
\caption{Upper panel: Normalized time-correlated PL decay at a
cavity length of $30~\mu$m (black), experimental APD instrument
response function (grey), and model fit with three exponential
decay channels (red solid line with $\tau_1=4.3$~ns, $\tau_2=33$~ns,
$\tau_3=660$~ns). Lower panel: Residuum of the fit.} \label{sfig1}
\end{figure}

Time-correlated photoluminescence (TCPL) decay traces recorded
with two different avalanche photodiodes (APDs) used in confocal
spectroscopy and cavity setups with 900 and 440 ps response times,
respectively, were modeled as a convolution of the APD instrument
response function (IRF) with multi-exponential decay functions as
follows:
\begin{equation}
I(t) =   I_0 + A_0 \cdot e^{-2 \cdot \left[(t-t_{0})/ w \right]^2}
+  \sum^{N}_{k=1} A_k \cdot e^{ \left(-t/\tau_k \right) },
\label{deceq}
\end{equation}
where the first term $I_0$ quantifies the APD dark counts, the
second term is the APD IRF approximated by a Gaussian with the
temporal resolution $w$, and the third is the sum of $N$
individual exponential decay channels $k$ with amplitude $A_k$ and
characteristic decay time $\tau_k$. The dark counts and the
response times $w$ of both APDs were calibrated experimentally as
in Fig.~\ref{sfig1}. The time $t_0$, set to the maximum of each
TCPL trace, was an input parameter to the fits with the amplitudes
$A_k$ and the decay times $\tau_k$ as free fit parameters. A
representative model fit to a TCPL trace (recorded for a cavity
length of $30~\mu$m)  obtained with there decay channels is shown
in Fig.~\ref{sfig1}.

\subsubsection{Determining the minimum number of decay channels}

\begin{figure}[h]
\includegraphics[scale=1]{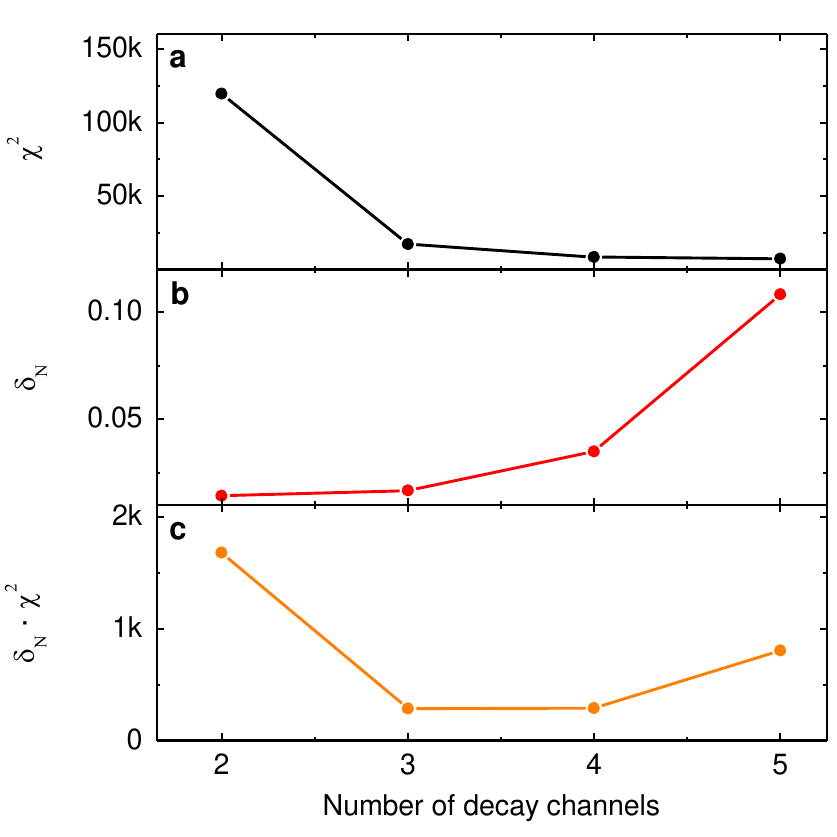}
\caption{\textbf{a}, Best-fit $\chi^2$ average, obtained by
averaging $\chi^2$ values of multi-channel best-fits for all
measurements at variable cavity lengths, as a function of the
number of decay channels. \textbf{b}, Averaged and normalized
parameter errors $\delta_N$ for the corresponding data.
\textbf{c}, Product $\delta_N \cdot \chi^2$ of both error types.
The minimum of the product at $N=3$ indicates that three decay
channels are best suited to approximate the multi-channel decay
characteristics of the HBL peak in Fig.~1 of the main text.}
\label{sfig2}
\end{figure}

To identify the minimum number of channels required to describe
the multi-exponential decay characteristics of the HBL emission,
the TCPL traces of the cavity measurements were fitted with a
varying number of channels using Eq. \ref{deceq}. The number of
possible channels was increased from two up to five. For each
number this analysis was applied to the whole set of TCPL
measurements from 35 $\mu$m down to 5 $\mu$m cavity length. The
quality of each fitting procedure was extracted by calculating the
$\chi^2$-value. An average $\chi^2$-value for all analyzed cavity
lengths is shown  in Fig.~\ref{sfig2}a. For an increasing number
of possible channels the $\chi^2$-value decreases down to the
measurement noise level.

Analyzing the $\chi^2$-value restricts the analysis to the quantity of
 the overlap of the best-fit function with the measured data, neglecting the possible errors of the
individual free fit parameters. According to Equation \ref{deceq}
each fitting procedure has a whole set of free fit parameters
with characteristic errors. For an adequate description of
the TCPL data these errors should be minimized. Therefore, an averaged error was
calculated for each fit using the errors of the individual free fit parameters.
A mean value for all corresponding lifetime traces results in
an overall error of the free fit parameters, $\delta_N$. The corresponding
errors are shown in Fig.~\ref{sfig2}b. The more possible decay channels contribute,
the higher is the overall error $\delta_N$ of the free fit parameters.

In order to have a quantity that respects both
error-types, the product $\delta_N \cdot \chi^2$ was calculated.  The
result is shown in Fig.~\ref{sfig2}c. This product has its
minimum for three decay channels, indicating
that this description is best suited to approximate the HBL emission
for analysis of both confocal and cavity TCPL data.

\subsection{Cavity-emitter coupling}

The general dependence of the cavity decay rate $\kappa$
and the coupling rate $g$ on the cavity length $L$  is given in the
framework of a coupled quantum well to a two
dimensional cavity as \cite{Savona1995}:
\begin{equation}
\kappa = 2 \cdot \frac{1-\sqrt{R}}{\sqrt{R}} \frac{c}{n_c L_c}.
\label{e1}
\end{equation}
The cavity decay rate $\kappa$ is extracted as the full-width at half-maximum
linewidth of broadband transmission spectra recorded for an empty cavity
(i.e. off MoSe$_2$-WSe$_2$ flakes) at a given
cavity length. The numbers of $\kappa$ we extract are consistent with
reflectivity values of $R=0.87$, which allows to simplify Eq.~\ref{e1} as:
\begin{equation}
\kappa(L)=\kappa_0 \cdot \frac{\lambda}{2L},
\label{eq:s1}
\end{equation}
where $\kappa_0$ is the cavity decay rate at a mirror separation
of $\lambda/2$. For our cavity system we obtain $\kappa_0 = 410$~meV.
Similar considerations for the collective coupling rate $g_k$
yield for each individual channel:
\begin{equation}
g_k(L)=g_{0,k} \cdot \sqrt{ \frac{\lambda}{2L}}. \label{eq:s2}
\end{equation}
In the next step we use $\gamma_{c}/\gamma_{fs} = 1 + C$, the ratio of the total decay rate in the
cavity system $\gamma_{c}$ to the free-space decay rate
$\gamma_{fs}$, together  with
the expression for the generalized Purcell factor $C = (4g^2/\gamma_{fs})/
(\kappa +\gamma_{fs} + \gamma_{d})$
\cite{Auffeves2010} to obtain the equation
for the individual decay channels:
\begin{equation*}
\gamma_{c,k} = \gamma_{fs, k} \cdot \left( 1 +
\frac{4g^2_k/\gamma_{fs,k}}{\kappa + \gamma_{fs,k} +
\gamma_{d}} \right),
\end{equation*}
with $\gamma_{d}$ being the dephasing rate of the emitter. The
functional dependence of the rate enhancement on the cavity length
can be obtained by using Eq. \ref{eq:s1} and \ref{eq:s2}. In
a final step the rate enhancement is converted into a lifetime
change via $\tau_{c,k} = 1/\gamma_{c,k}$ and
$\widetilde{\tau}_{k} = 1/\gamma_{fs,k}$. The resulting fitting
function takes $g_{0,k}$ and $\widetilde{\tau}_{k}$ as free fitting
parameters.


\end{document}